\documentclass{Interspeech}
\usepackage{booktabs}
\usepackage{multirow} 
\usepackage{float}
\usepackage[utf8]{inputenc}
\usepackage[table]{xcolor}
\usepackage[normalem]{ulem}

\interspeechcameraready

\newcommand{\benchmark}{\textsc{Sruti}}
\title{Recognizing Every Voice: Towards Inclusive ASR for Rural Bhojpuri Women}
\author[affiliation={1}]{Sakshi}{Joshi}
\author[affiliation={1}]{Eldho}{Ittan George}
\author[affiliation={1}]{Tahir}{Javed}
\author[affiliation={1}]{Kaushal}{Bhogale}
\author[affiliation={1}]{Nikhil}{Narasimhan}
\author[affiliation={1}]{\\Mitesh}{M. Khapra}
\affiliation{}{AI4Bharat, Indian Institute of Technology Madras}{India}
\email{sakshijcom@gmail.com, miteshk@cse.iitm.ac.in}

\begin{document}
\maketitle
\begin{abstract}
Digital inclusion remains a challenge for marginalized communities, especially rural women in low-resource language regions like Bhojpuri. Voice-based access to agricultural services, financial transactions, government schemes, and healthcare is vital for their empowerment, yet existing ASR systems for this group remain largely untested. To address this gap, we create \benchmark \footnote{\textit{\benchmark~ in Sanskrit means “that which is heard”}}, a benchmark consisting of rural Bhojpuri women speakers. Evaluation of current ASR models on \benchmark{} shows poor performance due to data scarcity, which is difficult to overcome due to social and cultural barriers that hinder large-scale data collection. To overcome this, we propose generating synthetic speech using just 25–30 seconds of audio per speaker from approximately 100 rural women. Augmenting existing datasets with this synthetic data achieves an improvement of 4.7 WER, providing a scalable, minimally intrusive solution to enhance ASR and promote digital inclusion in low-resource language.

%Digital inclusion remains a challenge for marginalized communities, especially rural women in low-resource language regions such as Bhojpuri. Voice-based access to agricultural services, financial transactions, government schemes, and healthcare is vital for their empowerment, yet existing ASR systems for this group remain largely untested. To address this gap, we create \benchmark \footnote{\textit{Sruti in Sanskrit means “that which is heard”}}, a benchmark containing rural Bhojpuri speakers. We evaluate current ASR models perform poorly on \benchmark, due to data scarcity. This data scarcity for this group is hard to overcome due to social and cultural barriers which hinder large-scale data collection. To overcome these challenges, we propose generating synthetic speech using only \remark{25–30} seconds of audio per speaker from approximately 100 rural women yielding just \remark{40–45 minutes} of training data. Augmenting existing datasets with this synthetic data achieves an improvement in WER by 4, offering a scalable, minimally intrusive solution to enhance ASR and promote digital inclusion in low-resource languages.
\end{abstract}

\section{Introduction}
Automatic Speech Recognition (ASR) is a critical technology for digital inclusion, especially for populations with low literacy who rely on speech to access services. In India, where 68.91\% of the population resides in rural areas, rural women face significant barriers to digital participation.\footnote{Census of India 2011, page 5 \url{https://censusindia.gov.in/nada/index.php/catalog/42617}}
ASR can bridge this gap by enabling access to government schemes, healthcare, and financial services. However, it is unclear whether current ASR systems work well for this demographic. Despite advancements driven by several large-scale data collection efforts in India, \cite{vaani2025,javed2024indicvoicesbuildinginclusivemultilingual,javed2022indicsuperbspeechprocessinguniversal,r2023springinxmultilingualindianlanguage,bhogale2022effectivenessminingaudiotext,OpenSpeechEkStep_ULCA_ASR}, improvements are often concentrated in high-resource languages and urban populations (see Table \ref{tab:comparison_table}). Without focused efforts, ASR risks reinforcing existing inequalities rather than reducing them. In this work, we focus on Bhojpuri-speaking rural women, a group that remains underserved by current ASR technology. %Addressing their needs requires direct intervention through targeted data collection, benchmarking, and model development.%Bhojpuri-speaking rural women represent one such group that is underserved by current ASR technology. Addressing their needs requires direct intervention through targeted data collection, benchmarking, and model development.

We first create a benchmark dataset for Bhojpuri-speaking women, covering key domains such as healthcare, finance, digital transactions, and government schemes which are important for digital inclusion. Data collection itself presents several hurdles, particularly in rural areas, which are defined as regions with fewer than 5,000 inhabitants, a population density below 400 per square kilometre, and where more than 25\% of the male workforce is engaged in agriculture\footnote{\url{https://www.india.gov.in/content/rural-indian}}. Trust is a major barrier, requiring collaboration with local intermediaries such as Accredited Social Health Activists (ASHA) workers and Auxiliary Nurse Midwives (ANMs), or NGOs who have established relationships within these communities. Additionally, social norms necessitate the involvement of women coordinators to facilitate participation. Another critical challenge is designing a questionnaire that ensures meaningful data collection while being mindful of rural contexts and community interests. Even after data collection, transcription remains a challenge due to the lack of standardized guidelines and the scarcity of skilled transcribers for Bhojpuri. %The language itself varies significantly across the northern belt of India, requiring careful listening to ensure accuracy. 
Despite these challenges, we collected 64.8 hours of speech, of which 17 hours has been transcribed, with the rest in the transcription pipeline.
% \begin{figure}
%     \centering
%     \includegraphics[width=\linewidth]{intro_spider.png}
%     \caption{Comparison of the number of hours in Hindi and Bhojpuri in 6 popular datasets for Indian languages.} 
%     \label{fig:size-comparison}
% \end{figure}

\begingroup
\setlength{\tabcolsep}{2pt} % Default value: 6pt
\begin{table}[t]
    \scriptsize
    \centering
    \caption{Number of transcribed hours in Hindi and Bhojpuri in 6 publicly available datasets for Indian languages}
    \label{tab:comparison_table}
    \begin{tabular}{l|cccccc}
    \toprule
    \textbf{Language} & \textbf{Vaani} & \textbf{IndicVoices} & \textbf{Kathbath} & \textbf{Shrutilipi} & \textbf{SpringINX} & \textbf{ULCA} \\
    \midrule
    Hindi    & 369  & 376  & 150  & 1620  & 351  & 2399 \\
    Bhojpuri & 23   & 0    & 0    & 0     & 0    & 60 \\
    \bottomrule
    \end{tabular}
\end{table}

\endgroup
% \begin{table}[t]
%     \centering
%     \caption{Number of transcribed hours in Hindi and Bhojpuri in 6 publicly available datasets for Indian languages}
%     \label{tab:comparison_table}
%     \begin{tabular}{lcc}
%     \toprule
%     \textbf{Dataset} & \textbf{Hindi} & \textbf{Bhojpuri} \\
%     \midrule
%     Vaani\cite{vaani2025}         & 367  & 22 \\
%     IndicVoices\cite{javed2024indicvoicesbuildinginclusivemultilingual}  & 333  & 0  \\
%     Kathbath\cite{javed2022indicsuperbspeechprocessinguniversal}     & 150  & 0  \\
%     Shrutilipi\cite{bhogale2022effectivenessminingaudiotext}   & 1620 & 0  \\
%     SpringINX \cite{r2023springinxmultilingualindianlanguage}   & 316  & 0    \\
%     ULCA        & 2399 & 60 \\
%     \hline
%     \end{tabular}
% \end{table}

While the collected data includes everyday conversations, responses to general icebreaker questions, and some read speech, our primary focus is on speech relevant to four critical domains for digital inclusion: \textit{health}, \textit{agriculture}, \textit{governance}, and \textit{finance}. Specifically, using the transcribed data, we develop {\benchmark,} a benchmark dataset of one hour, covering these four domains. %We evaluate existing ASR models and find that they perform poorly on this dataset. To improve performance, 
We then train an ASR model using Bhojpuri data from multiple existing sources, supplemented with data from Hindi, a related language with significant vocabulary overlap. However, even with these efforts, the Word Error Rate (WER) remains high at 33.3\%, compared to just 15\% for Hindi \cite{javed2024indicvoicesbuildinginclusivemultilingual}. This underscores the severe limitations of current ASR systems for low-resource languages and highlights the need for alternative solutions. %Without better data and improved modeling strategies, ASR will remain ineffective for Bhojpuri-speaking rural women, leaving them excluded from critical digital services.

Given the difficulty of collecting data from this demographic, can we rely on synthetic data to bridge the gap? Even this requires some seed data, but social barriers, household responsibilities, and limited mobility prevent rural women from participating in long recording sessions. 
%Ironically, the group that needs voice technology the most is also the least able to participate in large-scale data collection efforts. Social barriers, household responsibilities, and limited mobility make it difficult for rural women to spend hours contributing voice samples. 
Recognizing this, we investigate whether small voice samples, say 25-30 seconds from 100 women, can be leveraged to generate a large synthetic dataset. We utilize a multilingual speech synthesis model trained on a diverse set of Indian languages, though Bhojpuri itself was not included in the training data due to the unavailability of data. We rely on the assumption that the model's exposure to related languages, such as Hindi, will allow for meaningful transfer. Using just 39.4 minutes of real speech from 100 women, we generate 100 hours of synthetic Bhojpuri speech. This approach leads to a 4.7 point improvement in WER, demonstrating that synthetic data augmentation can significantly enhance ASR for low-resource languages as long as speech synthesis models trained on related languages are available. This scalable, practical approach offers a promising way forward for marginalized communities, where traditional data collection is infeasible. To encourage further research, we released\footnote{ \url{https://github.com/AI4Bharat/Sruti}} our dataset and code, advocating for similar efforts in other underserved languages.

\section{Related Work}
\noindent\textbf{Indic Datasets.} Currently, there are several large-scale dataset collection efforts for Indian languages, such as, Vaani \cite{vaani2025}, SpringINX \cite{r2023springinxmultilingualindianlanguage}, IndicVoices \cite{javed2024indicvoicesbuildinginclusivemultilingual}, Kathbath \cite{javed2022indicsuperbspeechprocessinguniversal} and ULCA \cite{OpenSpeechEkStep_ULCA_ASR}, along with web-mined datasets such as Shruitlipi \cite{bhogale2022effectivenessminingaudiotext}. However, as shown in Table~\ref{tab:comparison_table}, Bhojpuri remains severely underrepresented in these efforts. This %lack of coverage
poses significant challenges for training and evaluating ASR systems for marginalized communities speaking Bhojpuri, %low-resource languages,
a gap we aim to bridge in this work.

\noindent\textbf{Synthetic Data Generation.} %Synthetic data augmentation is a promising solution to data scarcity. 
Studies \cite{Joshi_2022,huang2023textgenerationspeechsynthesis} show that synthetic speech from advanced TTS models improves WER by supplementing limited training data. These methods use text-only corpora to generate training samples for fine-tuning ASR models. In low-resource settings, including dialects and minority languages, recent work \cite{yang2024enhancinglowresourceasrversatile} demonstrates that synthetic data reduces domain mismatch and improves robustness of models. Further, research on cross-lingual zero-shot multi-speaker TTS and voice conversion \cite{casanova2023asrdataaugmentationlowresource} demonstrates that even a single target-language speaker can significantly enhance ASR performance. These approaches %narrow the gap between models trained on synthetic and human speech, highlighting
show the potential of synthetic data in improving ASR, an approach that we also explore in this work. %These approaches effectively reduce the gap between models trained on synthetic and human speech, underscoring the role of synthetic data in improving accessibility to high-quality ASR systems, an approach that even we explore in this work.

\section{{\benchmark}: A Benchmark for Rural Bhojpuri Women}

In this section, we describe our data collection effort. %the development of the benchmark and training data used for augmenting our dataset. We begin by explaining how we prepared text prompts for users recording in Bhojpuri and how we collected text for generating synthetic data. Additionally, we discuss the challenges faced during field data collection in rural areas and the solutions we implemented to address them. 

\subsection{Selection of Target domains}
To understand the specific needs of rural women, we began by engaging with local communities through village visits. Through these interactions, we identified four key domains that are critical to enhancing digital inclusion for rural women: \textit{agriculture}, \textit{health}, \textit{government schemes}, and \textit{finance}. Agriculture is vital as women play a central role in farm management, and access to information on best practices and market prices can improve productivity. Health is another key domain, as rural women are often responsible for family health and need access to information on maternal care, disease prevention, and nutrition. Government schemes provide social welfare support, but limited awareness and accessibility prevent many women from benefiting fully. In finance, women in rural areas often lack access to essential services, making it difficult to engage with banking, loans, and insurance. These four domains were chosen for their direct impact on the lives of rural women.

%To create a dataset that truly reflects the experiences of rural women, we recognized that existing questionnaires, such as those in IndicVoices \cite{javed2024indicvoicesbuildinginclusivemultilingual}, lack rural-specific content, especially for extempore speech. The topics designed for urban populations do not effectively capture the daily lives and interests of rural women. Through our field visits, we observed that rural women commonly engage in activities like watching TV serials, participating in religious events, utilizing beauty parlour services, and engaging in home-based activities such as knitting. Based on these observations, we curated new target domains and categories that better reflect their lifestyles and interests.

%Our key observations indicated that the questions must be simple and engaging, as many participants have limited literacy. Conversations typically revolve around daily routines, social gatherings, and community events, rather than detailed discussions on agriculture, finance, tourism etc. Therefore, we introduced categories such as beauty, TV serials, home decor, and local functions and festivals to encourage richer and more natural responses. This tailored approach to domain selection aims to elicit vocabulary and expressions that are more representative of the target demographic, ultimately leading to a more engaging and authentic speech dataset.

\subsection{Demographic Considerations}
Bhojpuri is natively spoken in at least five states across India, but in this study, we focus on Uttar Pradesh, a northern state with a significant rural population. Specifically, we collect data from rural areas of three districts in Uttar Pradesh, \textit{viz.},  Bhadohi, Jaunpur, and Mirzapur. These villages have a population of less than 3,000, with variations in accent and vocabulary across the three districts. %regions represent diverse rural environments where Bhojpuri is predominantly spoken.
To ensure demographic diversity, we include women from four distinct age groups: 18-30, 30-45, 45-60, and 60+, representing a range of educational backgrounds, from those with formal schooling to those with little or no formal education. This helps us capture language use and communication styles from different age groups and education levels.

% This approach allows us to capture the linguistic variety and communication styles found across different age groups and educational backgrounds, ensuring the data represents the rural Bhojpuri-speaking population in its entirety.
\subsection{Designing Prompts for Data Collection}
%In our data collection process, we encountered a challenge similar to the one faced by IndicVoices, where 

Past data collection efforts \cite{javed2024indicvoicesbuildinginclusivemultilingual} have observed that simply asking participants to talk freely leads to very basic conversations, with short responses revolving around personal interests and lifestyle. To avoid this, we crafted targeted prompts to encourage more specific and meaningful responses. While we focused on key domains like health, agriculture, governance, and finance, we first ensured that the participants feel comfortable and engaged. To achieve this, we introduced additional domains such as beauty, home decor, entertainment, and local festivals as ice-breakers. These topics helped participants open up, and once they were comfortable, we could shift the focus to more specific tasks, such as answering questions or enacting scenarios related to the target domains (e.g., inquiring about a government scheme or inquiring about digital payments). We collected both read speech to ensure comprehensive vocabulary coverage and extempore speech to capture natural, conversational-style data. For the read speech component, we extracted sentences from the website of the Ministry of Rural Development, which offers comprehensive information on government schemes. In the health domain, we compiled texts related to women’s health initiatives and questionnaires commonly used at primary health centres (PHCs). Furthermore, we converted frequently asked questions (FAQs) from official government websites into %transactional text, making them suitable for use as 
transactional commands. To collect extempore data, we crafted simple questions related to these domains that rural women could easily relate to. These sentences were carefully translated into Bhojpuri by native speakers, creating a comprehensive set of 11746 transactional commands for read speech and 273 prompts to elicit extempore responses.

\subsection{Community Engagement and Awareness}
Obtaining informed consent and maintaining trust in rural communities is an essential aspect of our data collection process. Given the low level of digital literacy in these areas, ensuring proper consent while recording data presents a challenge. To address this, we collaborated with local ASHA workers and ANMs, trusted community health workers who regularly interact with rural women and are familiar with digital tools. We conducted awareness sessions to explain the purpose of our research, the role of speech technology in digital inclusion, the compensation they would receive and how the collected data would be securely handled. This approach ensured that the participants %and ensured they 
understood the benefits and implications of sharing their voice data. By engaging trusted local figures and providing clear information, we were able to secure informed consent, collect data ethically, and build trust, empowering participants to contribute to the development of inclusive technology.

%Obtaining informed consent and maintaining trust in rural communities is a critical part of our data collection process. Given the low level of digital literacy in these areas, recording data with proper consent can be challenging. To address this, we partnered with local ASHA workers and ANMs trusted community health workers who interact daily with rural women and are familiar with digital tools. We conducted educational sessions to explain the purpose of our research, the importance of speech technology for digital inclusion, and how the collected data will be securely handled and used.
%This approach not only helped to demystify the technology but also ensured that participants understood the benefits and implications of sharing their voice data. By engaging with the community through trusted local figures and providing clear, accessible information, we were able to obtain informed consent and collect data ethically, while fostering trust and empowering the participants to contribute to the development of inclusive technology.

\subsection{On-Field Data Collection}

We uploaded the transactional commands to be read and the extempore prompts onto the open-source Kathbath app, developed as part of the IndicVoices\cite{javed2024indicvoicesbuildinginclusivemultilingual} project for speech data collection. Once a participant gave informed consent, she was given credentials to log into the app. Upon logging in, participants could view the assigned tasks (commands and prompts) and record their responses by pressing the record button. To ensure smooth participation, a coordinator who is either our team member or a local ASHA worker or ANM, was present at all times to assist participants in navigating the app and addressing any technical difficulties. During the initial phase, we encountered significant challenges due to low connectivity in rural areas. Additionally, the limited availability of smartphones among participants necessitated data recording through a small number of coordinator devices. To address these challenges, we relocated data collection to government facilities such as primary health centres within the villages. These centres offered better connectivity and served as familiar and accessible locations for participants. While limited space at these centres introduced some background noise, this natural ambient sound reflects the rural environment and contributes to the authenticity of the dataset. Every recorded sample is verified by an in-house quality control expert. Only recordings where the spoken content remains clear and understandable, even in the presence of background noise or ambient chatter, are accepted. Additionally, only samples where responses are relevant to the given prompt are accepted. The participants and the quality control experts were compensated appropriately for their time.

%Data collection was conducted using the Kathbath app, with participants enrolled through a registration process that included obtaining informed consent. In the initial phase, we encountered significant challenges due to low connectivity in rural areas. Additionally, the scarcity of smartphones among participants meant that data recording had to be managed using one or two coordinator devices.
%To mitigate these issues, we relocated our data collection efforts to established government facilities, such as primary health centres, within the villages. These centers offered better connectivity and provided a convenient location for participants, as they are familiar and accessible within their community. While the limited space at these centres introduced some background noise, this natural ambient sound is an inherent part of the rural environment and contributes to the authenticity of our benchmark

\subsection{Transcription}
Transcription posed several challenges, as no established guidelines existed for Bhojpuri. To address this, we collaborated with Bhojpuri language experts to develop transcription guidelines, using Hindi transcription standards as a reference. Another major challenge was the lack of skilled transcribers, as Bhojpuri transcription is not a well-established industry. To overcome this, we conducted interviews and language assessments to evaluate candidates' proficiency in Bhojpuri spelling and grammar. To ensure high-quality transcripts for both benchmark and training data, we implemented a two-tier annotation process. At Level 1, an annotator performed the initial transcription, which was then reviewed and verified for accuracy at Level 2 by a second annotator. We engaged experienced Bhojpuri writers, including contributors to newspapers and other publications, as reviewers. This structured approach helped maintain accuracy and consistency. We observed that transcription speeds were significantly slow, as reviewers often had to listen to the audio multiple times to fully grasp the local dialect and vocabulary. Additionally, due to the lack of online resources for Bhojpuri, transcribers frequently needed to consult each other to confirm spellings. As a result, a transcriber typically produced only 20 minutes of transcribed data in a 6-8 hour workday. Similarly, reviewers required 6-8 hours to verify 20-40 minutes of transcriptions, depending on the complexity of the content. Due to the slow transcription process, we prioritized transcribing approximately one hour of speech from 51 speakers, focusing on content from the four target domains to serve as a benchmark. Additionally, to facilitate synthetic data generation, we transcribed 25-30 seconds of speech from 100 speakers. These short samples were used as prompts for synthesizing additional data in their voices as described later in Section \ref{sec:synthetic_data}.
\subsection{Data statistics}
\begin{figure}[H] 
    \centering
    \includegraphics[width=0.9\linewidth]{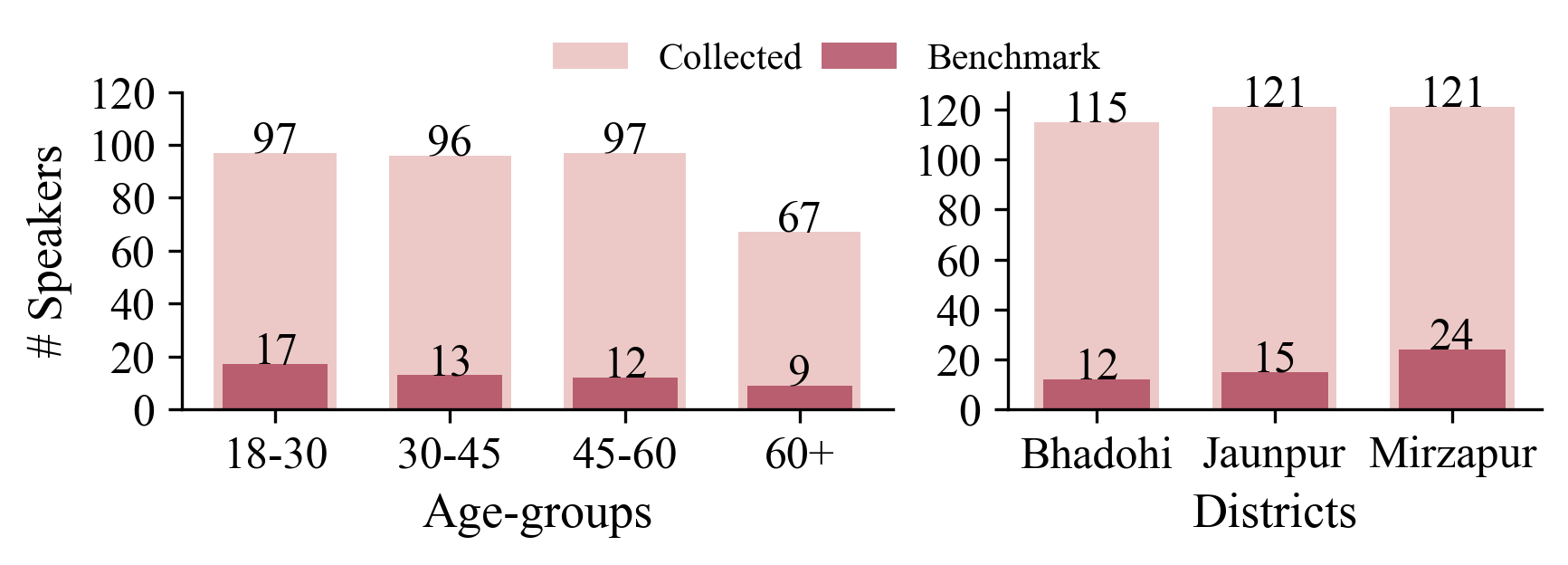}
    \caption{Distribution of number of speakers across different age groups \textbf{(left)} and different district \textbf{(right)}}
    \label{fig:age_and_district}
\end{figure}

\begin{figure}[h]
    \centering
    \includegraphics[width=0.9\linewidth]{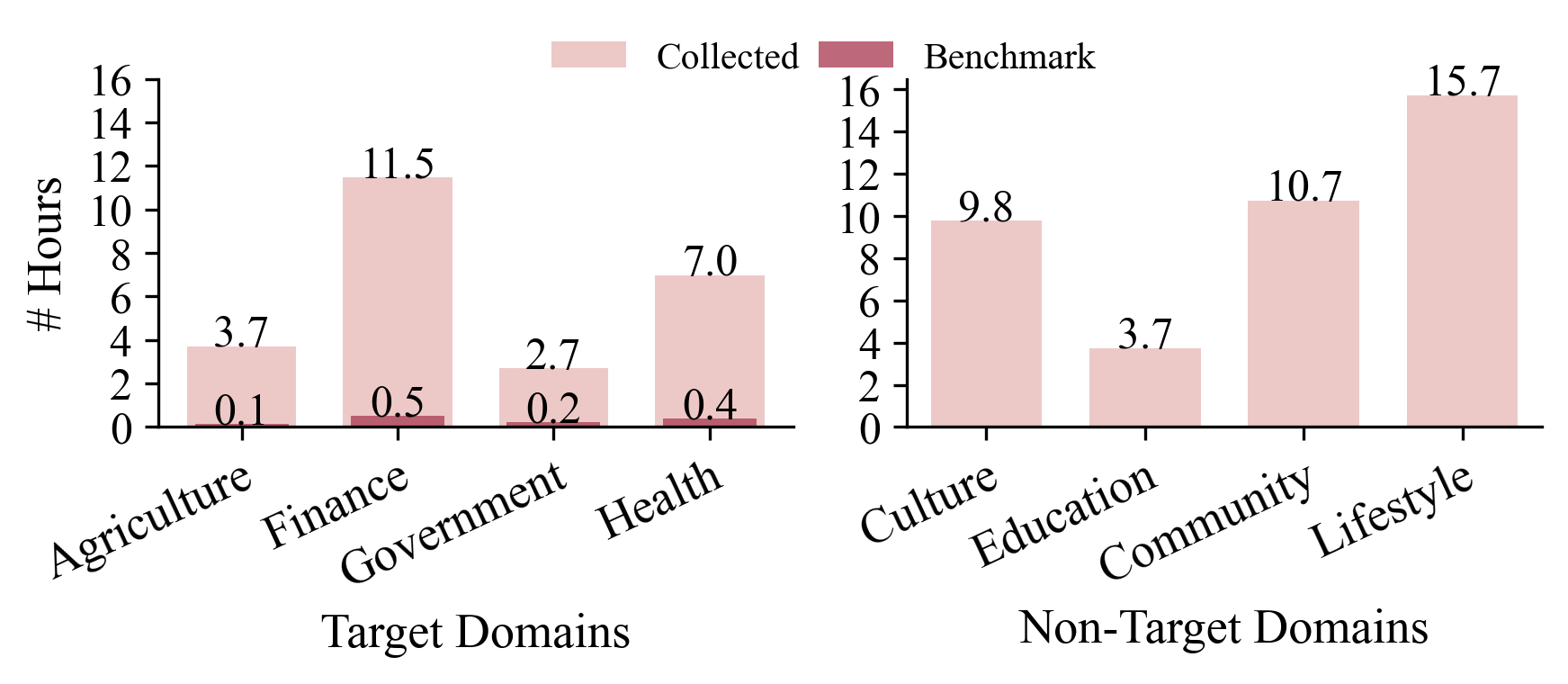}
    \caption{Number of hours of data in target domains \textbf{(left)} and non-target domains \textbf{(right)} in \benchmark}
    \label{fig:domains}   
\end{figure}

We collected a total of 64.8 hours of data from 277 speakers. Of these, we transcribe 72 minutes to create a benchmark with a total of 444 utterances from 51 speakers, covering the 4 target domains. In addition, we transcribed 1 transactional command and 1 short extempore response for 100 other speakers (who are not a part of the benchmark) amounting to a total of 39.4 minutes. Figure \ref{fig:age_and_district} (left) shows the distribution of speakers across age groups in the total data and the benchmark data. Figure \ref{fig:age_and_district} (right) shows the distribution of speakers across different districts. Figure \ref{fig:domains} shows the distribution of data across different domains. Apart from this, an additional 15.1 hours of data has also been transcribed but we could not use it for training as it contains data from the same speakers as in the benchmark.


\section{Synthetic Data Generation}
\label{sec:synthetic_data}
Since it is relatively easy to collect short audio samples from multiple speakers, we explored whether this data could be used to generate additional speech samples in the same voices. To achieve this, we employed a multilingual prompt-based speech synthesis model trained on 11 Indian languages \cite{anonymous2025}. It uses the English F5 model \cite{chen-etal-2024-f5tts} and fine-tunes it on 1417 hours of TTS summed up across 11 Indian languages. This model does not use any training data for Bhojpuri. We thus use it in a zero-shot setting hoping for transfer from related languages like Hindi seen in the training data. The model takes as input a short speech prompt from a given speaker along with its transcript and a new text that needs to be synthesised. For speech prompts we used the short audio samples that we collected from each of the 100 speakers. Along with this speech prompt, we input a Bhojpuri text sample, prompting the model to synthesize this text in the speaker's voice. Obtaining suitable Bhojpuri text for synthesis posed a challenge due to the limited availability of online resources. To address this, we adopted a two-pronged approach: (1) scraping text from multiple online sources, including \cite{leong2022bloomlibrarymultimodaldatasets,nllbteam2022languageleftbehindscaling,ws-2018-nlp-similar,ojha2019english}, and (2) generating English sentences in a rural context across four target domains using GPT-4o. These generated sentences were then translated into Bhojpuri using Google Translate. While translation errors were expected, they were not a significant concern since the primary objective was to generate synthetic speech in Bhojpuri to enhance vocabulary coverage.
In total, we synthesized 100 hours of speech data. Of this, 32.5 hours were derived from text related to the target domains, while the remaining 67.5 hours were based on general-domain text scraped from the web. This data used to augmented existing training datasets for Bhojpuri as described below. Additionally, since Hindi is a closely related language to Bhojpuri and spoken in the same regions, we collected 25–30 seconds audio samples from 100 rural Hindi-speaking women in the same districts. These samples were used to generate 100 hours of synthetic Hindi data, which was then used to train a bilingual model as described below.
%\textcolor{orange}{Along with generating synthetic data for Bhojpuri, we also produced 100 hours of Hindi synthetic data using recordings from 100 speakers from the same rural regions. To create the synthetic text for these recordings, we leveraged available text resources related to our target domains \cite{leong2022bloomlibrarymultimodaldatasets,kunchukuttan2018iitbombayenglishhindiparallel,gala2023indictrans2highqualityaccessiblemachine,ai4bharat2024rasa}, thereby producing a consistent and representative 100-hour synthetic corpus for Hindi, which we subsequently incorporated into our multilingual training setup.}

\section{Experimental setup}
We trained four models using varying amounts of data, as described below. Each model utilized the Conformer-L architecture \cite{gulati2020conformerconvolutionaugmentedtransformerspeech} with 130M parameters and a hybrid CTC + RNN-T loss function \cite{HybridRnntCtc}.The models were trained for 100 epochs with a peak learning rate of 5e-4, using the Noam learning rate scheduler and Adam optimizer. Training was conducted with a batch size of 32 audio samples per GPU on 8 NVIDIA H100 GPUs.

\noindent\textbf{M1:} This is a monolingual Bhojpuri model trained using a total of 133.4 hours of data collated from existing sources: SpeeS-IA \cite{interspeech2022}, ULCA NewsOnAir \cite{OpenSpeechEkStep_ULCA_ASR}, Vaani \cite{vaani2025}, and LIMMITS \cite{syspin}. 

\noindent\textbf{M2:} Here, in addition to the Bhojpuri data used to train M1, we use 376 hours of Hindi data from IndicVoices. Since Bhojpuri and Hindi are related languages we expect this additional data to improve performance due to crosslingual transfer. 

\noindent\textbf{M3:} Here, in addition to the data used for training M2, we use 100 hours of synthetic data generated in Bhojpuri. 

%as described in section \ref{sec:synthetic_data}.
\noindent\textbf{M4:} Here, in addition to the data used for training M3, we use 100 hours of synthetic data generated in Hindi.

\section{Results and Analysis}
We now summarize the results of different models on \benchmark. 
\subsection{Performance on \benchmark}
\noindent\textbf{Lower WERs.} From Table \ref{tab:main_table}, we first observe that the overall WER numbers are relatively high for this underserved demographic and low-resource language (29+ WER even in our best set up). For context, the best reported WER for Hindi on the IndicVoices test set is 15\cite{javed2024indicvoicesbuildinginclusivemultilingual}. This discrepancy is notable, considering the linguistic similarities between Bhojpuri and Hindi, even when accounting for the fact that the IndicVoices benchmark for Hindi includes both urban and rural speakers.

\noindent\textbf{Bilingual model is marginally better.} Next, comparing \textbf{M1} and \textbf{M2} we observe that on average adding Hindi data helps with the performance marginally better on two domains and marginally poor on two other domains.

\noindent\textbf{Synthetic data helps.} Introducing synthetic data substantially boosts performance across all the domains, with larger gains obtained by adding the Hindi synthetic data. Specifically, compared to \textbf{M1}, we observe an improvement of 2.7 - 6.4 WER, an average gain of 4.7 points, and a maximum improvement of 6.4 points in the Health domain. We attribute the larger gains obtained by adding the Hindi synthetic data, to the better quality and diversity of the text sourced for Hindi from multiple rich sources \cite{leong2022bloomlibrarymultimodaldatasets,kunchukuttan2018iitbombayenglishhindiparallel,gala2023indictrans2highqualityaccessiblemachine,ai4bharat2024rasa}. 
\subsection{Impact of Text from Target Domains}
To assess the impact of selecting in-domain text for generating synthetic audio, we trained two versions of the M4 model. In the first version, 32.5 of the 100 hours of synthetic training data were generated using text from targeted domains (this version is reported as the default model in Table \ref{tab:main_table}), while in the second version, all 100 hours were generated from general-domain text. The model trained with in-domain data from targeted domains significantly outperformed the model trained on general-domain text, as shown in Figure \ref{fig:domain_and_speaker_ablation}. This emphasizes the importance of using domain-specific text for effective synthetic data generation.

%To study the effect of selection of indomain text data, we trained two versions of the M4 model. In one version, 30 out of the 100 hours of synthetic training data were sourced from targeted domains, while in the other, all 100 hours came from the general domain. The model trained with targeted-domain data performed significantly better than the one trained on general-domain data, as highlighted in the Spider plot \ref{spider}. This also highlight the critical importance of selecting appropriate text for synthetic data generation.
\subsection{Does the Number of Speakers in Synthetic Data Affect Model Performance?}
\begin{figure}[!t]
  \centering
  \includegraphics[width=\linewidth]{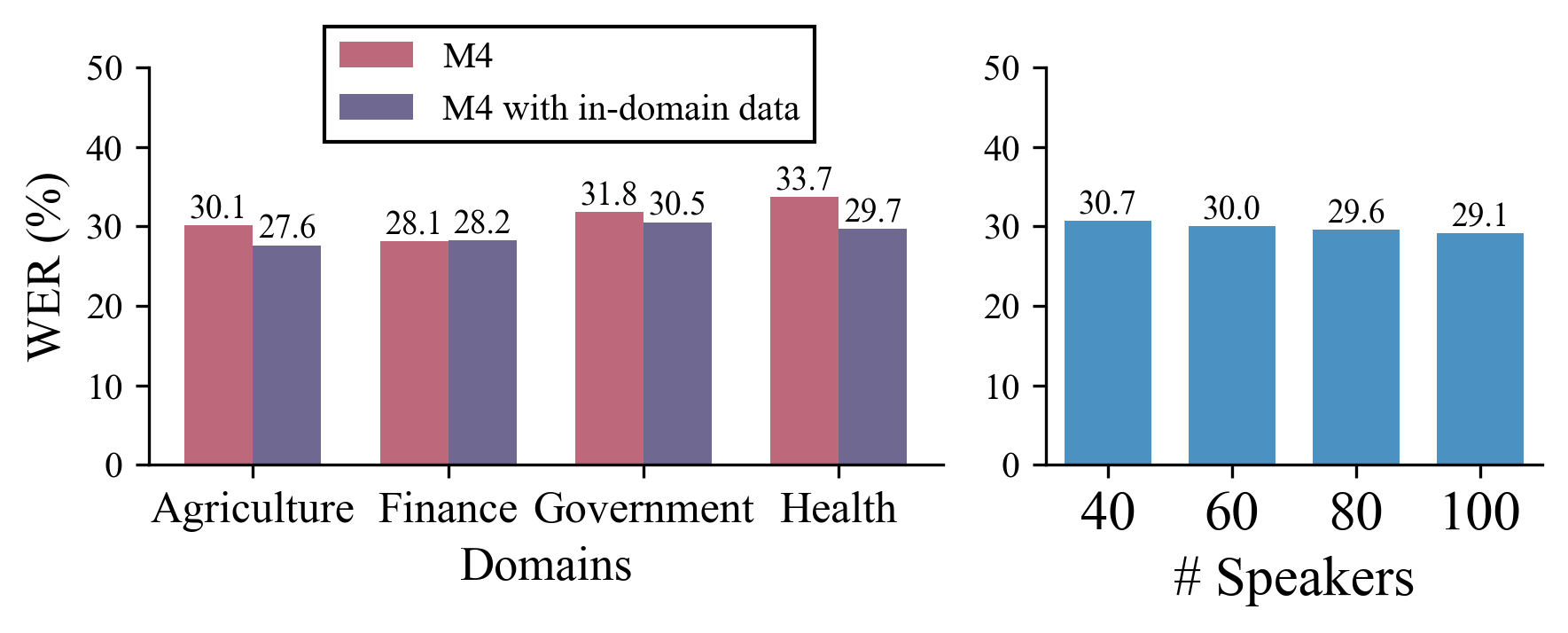}
  \caption{\textbf{(Left)} Model's performance with and without in-domain synthetic data. \textbf{(Right)} Effect of number of speakers in the synthetic data on model performance.}
  \label{fig:domain_and_speaker_ablation}
\end{figure}

To investigate the effect of the number of unique speakers in synthetic training data, we conducted an ablation study. In this study, the total audio duration and textual content were kept constant, while the number of unique speakers was varied (e.g., 40, 60, 80, 100). Specifically, we generated 100 hours of synthetic data using the same text as used for creating the synthetic data for the default M4 model, but with prompts from different numbers of speakers (40, 60, 80) for 
% both Hindi and 
Bhojpuri. Our results show that increasing the number of speakers generally leads to performance improvements (see Figure \ref{fig:domain_and_speaker_ablation}). However, this improvement may plateau unless the total amount of data is increased, which in our study is currently capped at 100 hours.

\begingroup
\setlength{\tabcolsep}{5pt} % Default value: 6pt
\begin{table}[!t]
    \centering
    \scriptsize
    \caption{Comparison of Model Performance Across Domains}
    \label{tab:main_table}
    \begin{tabular}{l|cccc|c}
        \toprule
         & \textbf{Agriculture} & \textbf{Finance} & \textbf{Government} & \textbf{Health} & \textbf{Average} \\
        \midrule
        \textbf{M1} & 30.3 & 32.4 & 36.8 & 36.1 & 33.8 \\ 
        \textbf{M2} & 31.0 & 33.4 & 34.4 & 34.7 & 33.3 \\ 
        \textbf{M3} & 26.4 & 34.4 & 32.9 & 32.5 & 31.5 \\ 
        \textbf{M4} & 27.6 & 28.2 & 30.5 & 29.7 & 29.1 \\ 
        \bottomrule
    \end{tabular}
\end{table}
\endgroup

\if 0
\begin{table}[htbp]
    \centering
    \caption{Word Error Rate (WER) by number of unique speakers in synthetic data, with total duration (100 hours) and text content held constant}
    \label{table:Number of speakers }
    \begin{tabular}{cc}
    \hline
    \textbf{Numbers of speakers} & \textbf{WER} \\
    \hline
    40  & 30.7 \\
    60  & 30.0 \\
    80  & 29.6 \\
    100 & 29.0 \\
    \hline
    \end{tabular}
   \end{table}
 \fi 
% \end{table}
% \begin{figure}[H]
%   \centering
%   \includegraphics[width=0.8\linewidth]{.png}
%   \caption{Effect of numbers of speakers in synthetic data}
%   \label{fig:Number of speakers }
% \end{figure}
\section{Conclusion}
In this work, we address the lack of ASR data and benchmarks for rural Bhojpuri-speaking women. Through field visits and community engagement, we overcome social barriers and collect data in four key domains, \textit{health}, \textit{agriculture}, \textit{governance}, and \textit{finance}, that are vital for digital inclusion. We discuss challenges in data collection process requiring the need for local intermediaries and the impact of social norms and logistical issues in rural areas. Using this benchmark, we evaluate existing ASR models and find them lacking. We show that adding data from a related language, like Hindi, along with synthetic data generation, improves performance. Since synthetic data only requires short voice samples, it offers a cost-effective and minimally intrusive solution for enhancing ASR in low-resource languages.

%In summary, our results indicate that synthetic data augmentation, particularly when combined with multilingual training, can substantially improve ASR performance for low-resource languages. However, optimizing the quality of synthetic text inputs remains an essential area for future work to reduce recognition errors specific to Bhojpuri further.

\section{Acknoweldgments}
We would like to thank the Bill \& Melinda Gates Foundation for their generous support in making this work possible. We gratefully acknowledge Yotta Infrastructure for providing access to GPU resources that enabled large-scale model training. We would also like to thank the EkStep Foundation and Nilekani Philanthropies for their generous grant, which supported the hiring of human resources and access to cloud infrastructure essential for this work. We are thankful to the Aripana Foundation for their on-ground support and coordination during data collection. We are deeply grateful to the ANMs and ASHAs who assisted us during the recording process in the field. We also extend our thanks to the team of AI4Bharat for their help in collecting data from native speakers of different languages across the country. We sincerely appreciate the dedicated efforts of our entire team throughout the project. Most importantly, we extend our heartfelt gratitude to every rural woman who participated in the recording efforts this work would not have been possible without them.
\bibliographystyle{IEEEtran}
\bibliography{mybib}

\end{document}